\documentclass[
    ,final            
  ]
  {aipproc}

\layoutstyle{8x11single}

\usepackage{amsmath}
\usepackage{amsfonts}
\usepackage{amssymb}
\usepackage{textcomp}
\usepackage{graphicx}
\usepackage{bm}
\usepackage{color}
\usepackage{verbatim}
\usepackage{subfig}

\def\beq{\begin{equation}}
\def\eeq{\end{equation}}
\def\beqn{\begin{eqnarray}}
\def\eeqn{\end{eqnarray}}
\newcommand{\be}{\begin{equation}}
\newcommand{\ee}{\end{equation}}
\newcommand{\ba}{\begin{eqnarray}}
\newcommand{\ea}{\end{eqnarray}}
\newcommand{\ben}{\begin{enumerate}}
\newcommand{\een}{\end{enumerate}}

\newcommand{\p}{\partial}
\newcommand{\ph}{\varphi}

\newcommand{\la}{\langle}
\newcommand{\ra}{\rangle}

\newcommand{\rar}{\rightarrow}

\begin{document}

\begin{flushright}
 ITEP-TH 04/14
\end{flushright}

\title{Baryon and chiral symmetry breaking \footnote{Contribution to the Proceedings of the 
II Russian-Spanish Congress ``
Particle and Nuclear Physics at all Scales and Cosmology'', 
Saint-Petersburg, October 1-4, 2013
}}


\author{A. Gorsky}{
  address={
Institute for Theoretical and Experimental Physics (ITEP), \\ Moscow, Russia
}
,altaddress={
Moscow Institute of Physics and Technology (MIPT),\\
Dolgoprudny, Russia
}
}

\author{A. Krikun}{
  altaddress={
Institute for Theoretical and Experimental Physics (ITEP), \\ Moscow, Russia
}
,address={
NORDITA, KTH Royal Institute of Technology and Stockholm University \\
Stockholm, Sweden
}
}

\begin{abstract}
 We briefly review the generalized Skyrmion model for the baryon recently suggested by  us.
It takes into account the tower of vector and axial mesons as well as the chiral symmetry breaking.
The generalized Skyrmion model provides the qualitative explanation of the Ioffe's formula for the baryon mass.
\end{abstract}

\maketitle


\section{Introduction}
In \cite{instanton1, instanton2} we have introduced the generalization of the Skyrmion model for the baryon. The
idea was to take into account the infinite tower of the vector and axial mesons and the chiral condensate simultaneously.
These data are most effectively packed into the 5-dimensional holographic model for QCD \cite{hard-wall} in the curved geometry.  The expansion of
5d gauge fields with $SU(N_f)_L\times SU(N_f)_R$ gauge group  into Kaluza-Klein modes yields the whole tower of the massive mesons while the chiral condensate is encoded in the boundary condition for the bifundamental scalar. The conventional Skyrmion is just the solitonic configuration built from the pions only however there were some attempts to include the vector mesons into the solution (see \cite{mei} for review). In particular the
stabilization of the Skyrmion via the $\omega$ - mesons has been suggested \cite{nappi1, nappi2} which substitutes the Skyrme term mechanism. More recently the Skyrmion has been discussed in 5d model in the flat space and it was found that the vector and axial mesons strongly influence the Skyrmion solution \cite{sutc}.
In \cite{instanton2} we found the generalization of the Skyrmion solution with highly nontrivial role of the whole tower of the axial and vector mesons. 

There was  longstanding puzzle concerning the role of the QCD chiral condensate in the baryon mass generation. This subject was triggered by
the sum rule calculation of the baryon mass  \cite{Ioffe}  resulted in the following surprising formula
\beq
\label{Form_Ioffe}
M^3 = -8 \pi^2 \la \bar{q}q \ra,
\eeq
which is valid with reasonable accuracy. It implies that the huge amount of the baryon mass is due to the chiral condensate.  It was suggested that the naive formula
\beq
\label{linear}
M_B= m_0 + \Delta( \la \bar{q}q \ra),
\eeq
where $m_0$ is condensate independent, well interpolates the physics. The similar formula has been
also used in the parity-doublet model and the Chiral Quark-Soliton model (see \cite{diak} for a review).

The conventional  Skyrmion was identified as the instanton solution at the 4d slice of the 5d AdS-like space \cite{Sakai-baryon} where the holographic RG coordinate plays the role of the Euclidean time direction. It can be also considered along the Atiyah-Manton picture  \cite{Atiah-Manton} as the
instanton trapped by the domain wall localized in the holographic coordinate. The holographic perspective suggests the proper generalization of the instanton solution which would take into account the massive mesons and chiral condensate. It is similar to the dyonic instanton solution found in \cite{lt} in the 5d SUSY gauge theory. In \cite{instanton1} it was shown that the solution of the dyonic instanton type exists at the particular choice of the boundary conditions. It has strong similarity with the Atiyah-Manton picture of the domain wall localized at the holographic coordinate.

 The   numerical solution to the equations of motion were found in \cite{instanton2} and it was demonstrated analytically that it carries one unit of the baryon charge. Its stabilization is dictated by the chiral condensate and the confinement scale. Remarkably our dyonic instanton solution yields the elegant solution to the longstanding puzzle concerning the origin of the  Ioffe's formula  (\ref{Form_Ioffe}). It turns out that the dependence of the baryon mass on the chiral condensate has two branches. At small values of the condensate the mass is condensate independent  while above  some critical value  it 
grows according to the Ioffe's formula.
The analysis shows that the nature somehow dynamically selects
the value of the condensate just at the intersection of two branches. Therefore we have the natural
explanation of the Ioffe's formula while at small values of the condensate
the baryon mass becomes almost condensate independent.

\section{Brane picture of baryon as dyonic instanton}

\begin{figure}[h!]
\includegraphics[width=0.5 \linewidth]{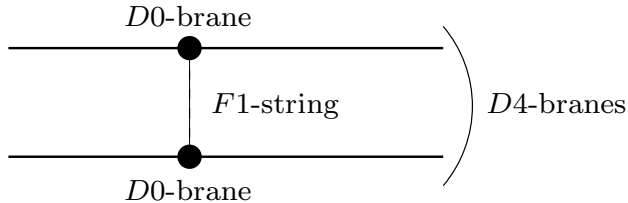}
\caption{\label{D4branes} Schematic picture of the dyonic instanton in 5d supersymmetric theory}
\end{figure}

The similarities between dyonic instanton in five-dimensional $\mathit{N}=1$ supersymmetric Yang-Mills theory and holographic QCD model become apparent if one examines the problem from the brane-picture point of view. One can consider the  supersymmetric Yang-Mills with the gauge group $SU(N)$ as a theory realized by the massless modes of open strings on the worldvolume of the stack of N D4-branes in IIB superstring theory. The breaking of the gauge symmetry is realized via separation of the branes in the transverse coordinate (see Fig. \ref{D4branes}). The value of this separation is in one-to-one correspondence with the the vacuum expectation value (VEV) of the scalar field which brakes the symmetry in the field-theoretic formulation. The gauge bosons associated with the broken symmetry acquire mass proportional to the separation scale. In this picture it was shown in \cite{hashimoto, tubes} that for the SU(2) gauge group the dyonic instanton is realized as a pair of D0 branes, which are located on the 
separated D4 branes, and an electrically charged F1 string, which is stretched 
between them. Being proportional to the length of the F1 string, the mass of the instanton is determined by the VEV of the scalar field. Furthermore, for topological charges larger then one this construction is found to be unstable to the formation of the tubular D2 brane out of F1 string. In this case the instanton assumes the form of the monopole ring, whose radius is stabilized by the angular momentum.

\begin{figure}[h!]
\includegraphics[width=0.5 \linewidth]{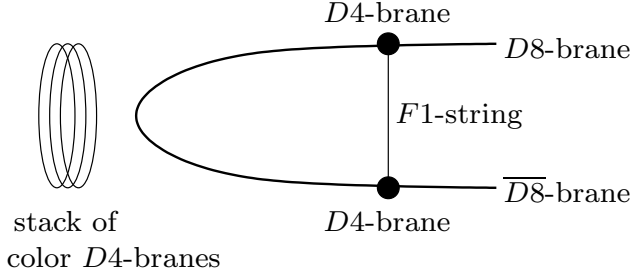}
\caption{\label{cigar} Schematic picture of the baryon is Sakai-Sugimoto model}
\end{figure}

The very similar setup can be found in the holographic models of QCD. In Sakai-Sugimoto model \cite{ss} the gauge sector of QCD is represented by the massless open string modes on the stack of $N_c$ D4 ``color'' branes, compactified in one spacial direction in order to break supersymmetry. The quarks are represented by the two  additional stacks of $N_f$ $D8$ and $\overline{D8}$ ``flavor'' branes. The embedding of these branes in the 10-dimensional space is described by their position in the compactified dimension and the distance to the stack of D4 ``color'' branes and has the cigar-shaped form (see Fig.\ref{cigar}), so that the $D8$ and $\overline{D8}$ branes join at particular distance from the ``color'' branes. As the Chan-Paton factors of the string connecting $D8$ and $\overline{D8}$ ``flavor'' branes belong to the fundamental and anti-fundamental representations of the flavor $SU(N_f)$ group, the chiral symmetry breaking operator $\bar{q}q$ is naturally described by this type of open string \cite{berg}
. From the 
point of view of the theory on the stack of ``color'' branes the Sakai-Sugimoto model describes $SU(N_c)$ nonsupersymmetric Yang-Mills theory with $N_f$ flavors of fundamental quarks and the spontaneous chiral symmetry breaking therein. This theory lives effectively on the four-dimensional intersection of the ``color'' and ``flavor'' branes. Integrating out the compactified directions of the bulk space and implementing the AdS/CFT duality one can obtain from this setup the usual chiral perturbation theory Lagrangian supplemented with the interactions of pions with the whole tower of vector and axial-vector mesons \cite{ss}.

In Sakai-Sugimoto model we find the situation similar to the one described above. The chiral symmetry is broken and the symmetry breaking scale $\la \bar{q} q \ra$ is defined by the separation of D8 branes. In the case of $N_f=2$ the pair of D4 branes can be formed in the worldvolume of the respected D8 branes, similarly to the D0 branes on the D4 stack in the case of dyonic instanton. These D4 branes fill the compactified directions of the bulk space and look as an instanton from the point of view of the theory on the stack of $N_c$ ``color'' branes \cite{witten1, witten2}. Furthermore, the instanton solutions, which possess nonzero topological charge, can be shown to describe baryons upon AdS/CFT duality \cite{Sakai-baryon}. The F1 string stretched between $D8$ and $\overline{D8}$ branes, which is dual to the $\la \bar{q} q \ra$ operator, is trapped by these instantons. This mechanism prevents the fall of the string to the tip of ``the cigar'' thus preventing its shrinking, which is the well-known problem 
in Sakai-Sugimoto model \cite{berg}. Therefore in our investigation the baryon physics described by the instantonic branes is tightly related to the chiral physics described by the F1 string and this relation provides the stabilization mechanism of the solution.
This leads us to the conclusion that the dyonic instanton and holographic baryon have much in common.

\section{Realization in the hard-wall holographic model}

To proceed further and construct the baryonic solution in the holographic QCD model taking the similarities mentioned above as a guideline we chose the simple five-dimensional ``hard-wall'' model \cite{hard-wall}. This model can be considered as a somewhat simplified result of the integration of all the dynamics of the Sakai-Sugimoto model along the compactified spacial directions, which leaves only lowest lying string modes propagating in 5-dimensional AdS-space with an additional cut off at particular scale $z_m$:
\begin{equation}
\label{metric}
ds^2 = \frac{1}{z^2} (dt^2 - dz^2 - dx_i^2), \quad i=1 \dots 3, \quad z<z_m
\end{equation}
The field content of the model includes two gauge fields $L_\mu^a$ and $R_\mu^a$ in the adjoint representation of the flavor group $U(N_f)$ dual to the left and right quark currents, respectively. These fields in the brane picture are identified with the lowest lying modes of the open stings, which have two ends attached to the flavor D8 branes. Their vector and axial linear combinations will describe the full towers of vector and axial mesons in the model. Another field is a bifundamental scalar $X^{\alpha \beta}$ which is dual to the quark bilinear operator $\bar{q}^\alpha q^\beta$. It is the tachionic mode of the string stretched between $D8$ and $\overline{D8}$ branes as discussed above. The vacuum expectation value of the operator $\bar{q}^\alpha q^\beta$ is a chiral condensate. Hence by the duality argument \cite{gubser-klebanov} the classical profile of the scalar field $X$ is proportional to this symmetry breaking scale. In what follows we take $N_f = 2$. The action reads as (we use the notation of \
cite{Krik1,Krik2,tensor})
\begin{equation}
\label{action}
S = \int d^3 x dt dz \bigg\{ \frac{1}{z} \left( - \frac{1}{4 g_5^2} \right) (F_L^2 + F_R^2)
+ \frac{\Lambda^2}{z^3} (D X)^2 + \frac{\Lambda^2}{z^5} \  3|X|^2 \bigg\},
\end{equation}
where the covariant derivative is $D_\mu X = \p_\mu X - i L_\mu X + i X R_\mu $. The constants in the action can be fixed by matching the two-point functions calculated in the holographic model to the leading terms in the sum rule approach \cite{hard-wall, Krik2}: $\Lambda^2 = \frac{3}{g_5^2}$, $\frac{1}{g_5^2} = \frac{N_c}{12 \pi^2}$. Apart of the Yang-Mills part of the action one needs to consider the Chern-Simons part, which is
\begin{equation}
\label{CS}
S_{CS} = \frac{N_c}{24 \pi^2} \int \frac{3}{2} \left \{\hat{L} \ tr (F_L \tilde{F_L}) - \hat{R}  \ tr ( F_L \tilde{F_L}) \right \}.
\end{equation}
Here we mark with a hat the Abelian (related to the Cartan subalgebra of $U(N_f)$) parts of the gauge fields. One should recall \cite{Sakai-baryon, Wulzer} that as the Abelian part of the gauge field is dual to the $U(1)$ quark current, the vector combination $\hat{L}_\mu + \hat{R}_\mu = \hat{V}_\mu$ is dual to the baryon number current. One can see from (\ref{CS}) that the source for the temporal component of the baryon current -- the baryon charge -- is
\begin{equation}
\label{baryon}
Q_B =\frac{1}{16 \pi^2} \int d^3 x dz \epsilon^{\mu \nu \lambda \rho} \left[F_L^{\mu \nu} F_L^{\lambda \rho} - F_R^{\mu \nu} F_R^{\lambda \rho} \right].
\end{equation}
This is identified with the difference between four-dimensional topological charges of the left and right nonabelian field configuration in the constant time slice. Thus in order to describe the baryon we need to find the solution to the equations of motion of the model which has nontrivial topological charge -- the instanton. 

In the following we use the ansatz similar to the Witten cylindric ansatz \cite{Witten_inst} with the holographic coordinate $z$ along the axis of the cylinder and the spacial radius $r$ in the slice of constant $z$. Now the model can be rewritten in terms of two-dimensional fields:
\begin{align}
\label{ansatz}
L_j^a &= -\frac{1 + \xi_2 (r,z) + \eta_2 (r,z)}{r} \epsilon_{jak} \frac{x_k}{r} 
+ \frac{\xi_1 (r,z) + \eta_1 (r,z)}{r} \left( \delta_{ja} - \frac{x_j x_a}{r^2} \right) 
+ \big(V_r(r,z) + A_r(r,z)\big)  \frac{x_j x_a}{r^2}, \\
\notag
R_j^a &= -\frac{1 - \xi_2 (r,z) + \eta_2 (r,z)}{r} \epsilon_{jak} \frac{x_k}{r} 
+ \frac{\xi_1 (r,z) - \eta_1 (r,z)}{r} \left( \delta_{ja} - \frac{x_j x_a}{r^2} \right) 
+ \big(V_r(r,z) - A_r(r,z) \big) \frac{x_j x_a}{r^2}, \\
\notag
L_5^a &= \big(V_z(r,z) + A_z(r,z)\big) \frac{x_a}{r}, \qquad
R_5^a = \big(V_z(r,z) - A_z(r,z)\big) \frac{x_a}{r},\\
\notag
X &= \chi_1(r,z) \frac{\mathbf{1}}{2} + i \chi_2(r,z) \frac{\tau^a x^a}{r},
\end{align}
where $\tau^a$ are the generators of $SU(2)$ obeying the commutation relation $[\tau^a \tau^b] = - i \epsilon^{abc} \tau^c$. Moreover, it is convenient to introduce the phases of the complex combinations of the scalar fields
\begin{align}
\label{phases}
\eta_1 &= \phi \cos(\theta) \cos(\alpha), & \eta_2 &= \phi \cos(\theta) \sin(\alpha), \\
\notag
\xi_1 &= \phi \sin(\theta) \cos(\alpha-\omega), & \xi_2 &= \phi \sin(\theta) \sin(\alpha-\omega), \\
\notag
\chi_1 &= \chi \cos(\gamma), & \chi_2 &= \chi \sin(\gamma).
\end{align}
Studying the equations of motion on the boundaries we find, that the equation for the modulus $\chi$ is always satisfied by
\begin{equation}
\label{chi}
\chi = m z + \sigma z^3.
\end{equation}
This is the consequence of the fact that the field $X$ is holographically dual to the quark bilinear operator $(\bar{q}q)$ \cite{hard-wall, gubser-klebanov}. The form of the solution is dictated by the canonical dimension $\Delta$ of this operator: one branch behaves as $z^{4-\Delta}$ and the other as $z^{\Delta}$. The two dimensional coefficients are related to the source of the operator (which is for  $(\bar{q}q)$ the quark mass $m$) and the vacuum expectation value $\la \bar{q}q \ra$, namely the quark condensate. The parameter $\sigma$ is related to the quark condensate as \cite{Krik1}
\begin{equation}
\label{sigma}
\sigma = \frac{N_f}{3 \Lambda^2}  \langle \bar{q}q \rangle \approx (460 Mev)^3
\end{equation}
In what follows we study the chiral limit $m = 0$. 

The energy functional of the model that we get by substituting \ref{ansatz} and \ref{phases} into the action \ref{action} contains several potential terms. One of them is
\begin{equation}
\label{potential1}
 \frac{1}{z r^2}  \Big[(\phi^2 - 1)^2 + \phi^4 \sin(2 \theta)^2 \cos(\omega)^2 \Big].
\end{equation}
Since this potential is positively definite, the lowest energy solution is defined by the two terms separately. The first leads to $\phi = 1$, the second -- to either $\theta = \frac{\pi}{2} n$ or $\omega = \frac{\pi}{2} + \pi m$ ($m,n \in \mathbb{Z}$). Next, the other potential term is
\begin{equation}
\label{potential2}
\frac{6}{z^3} \chi^2 \ph^2 \big[ \cos(\theta)^2 \ \cos(\gamma - \alpha)^2 
+ \sin(\theta)^2 \ \sin(\gamma  - \alpha + \omega))^2 \big].
\end{equation}
and consequently in the vacuum state either $\sin(2 \theta) = 0$ and
\begin{align*}
\gamma - \alpha = \frac{\pi}{2} + \pi n, & \quad n \in \mathbb{Z} &  \mbox{ for } &\sin(\theta) = 0, \\
\gamma-\alpha = -\omega   + \pi n, & \quad n \in \mathbb{Z} & \mbox{ for } &\cos(\theta) = 0,
\end{align*}
or  $\sin(2 \theta) \neq 0$ and
\begin{align*}
\omega = \frac{\pi}{2} + \pi n, \qquad n \in \mathbb{Z}.
\end{align*}
Hence the vacuum state of our model is described by two quantum numbers: $n$ and $m$, which identify the values of phases $\omega$ and $\theta$. In order to describe the solution with nonzero topological charge we need to choose different vacuum states at different boundaries of the 4D AdS slice, thus realizing the instanton-like tunneling process. 

One way of doing this is to consider constant phase $\theta = 0$ and singular behavior of the scalar modulus $\phi$ at the $r \rar \infty$ boundary \cite{instanton1}. In this scenario the scalar field vanishes at the point $z=z_0$ at asymptotical infinity, what allows for the kink-like behavior of the phase $\omega$ is $z$ direction. Hence the solution obtained in this way is a domain wall placed in the holographic direction parallel to the boundary of $AdS$. The field strength created on the boundary by the instanton trapped by this wall translates directly to the Skyrmion-like field configuration on the dual theory. The realization of the Skyrmion as an instanton on the domain wall was proposed in \cite{Atiah-Manton} and this solution generalizes this idea to the holographic QCD model.

The other way of realizing soliton in this model was elaborated in \cite{instanton2}. It is more general then the previous one and is free from the singular points. In order to get nonzero baryon charge without demanding for the vanishing of the scalar field one needs to consider simultaneous change of the phase $\omega$ along the $r$ direction and the phase $\theta$ along the $z$ direction. Taking the values presented in Table \ref{table1} one can show, that the baryon charge (\ref{baryon}) on this solution is defined by the boundary conditions and is equal to one.
\begin{table}[h!]
\begin{tabular}{|c|c|c|c|c|}
 \hline
$\quad r \quad$ & $0 \rar \infty$ & $\infty$ & $\infty \rar 0 $ & $ 0 $\\ \hline
$z $ & $ 0 $ & $ 0 \rar z_m $ & $ z_m $ & $ z_m \rar 0 $\\ \hline \hline
$\theta $ & $ 0 $ & $ 0 \rar \frac{\pi}{2} $ & $ \frac{\pi}{2} $ & $ \frac{\pi}{2} \rar 0$ \\ \hline
$\omega $ & $ -\frac{\pi}{2} \rar \frac{\pi}{2} $ & $ \frac{\pi}{2} $ & $ \frac{\pi}{2} \rar -\frac{\pi}{2} $ & $ -\frac{\pi}{2} $\\ \hline
$\gamma $ & $ \pi \rar 0 $ & $ 0 $ & $ 0 \rar \pi $ & $ \pi$ \\ \hline
\end{tabular}
\caption{\label{table1} The boundary values for the phases (\ref{phases}), which describe the solution with unit baryon charge. The phase $\alpha$ is set to $-\tfrac{\pi}{2}$ by residual gauge symmetry.}
\end{table}

One can also observe that the field $X$ in this solution behaves exactly the same as the chiral field $U$ in the ``hedgehog'' Skyrmion. 

\begin{figure}[h!]
\includegraphics[width=0.5 \linewidth]{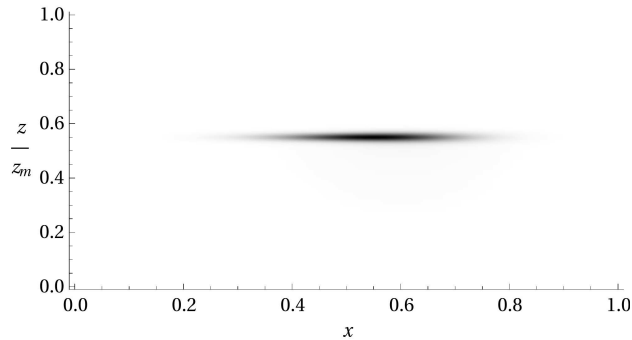}
\caption{\label{disk} The energy density of the solution with small $z$ radius and large $r$ radius. The $r$ coordinate is rescaled so that $r \rar \infty$ boundary is located at $x=1$.}
\end{figure}

The solution with asymptotic values of phases described above can be studied numerically. Interestingly enough the solution tend to shrink in the $z$ direction and expand in the $r$ direction. That means it looks more or less as a thin disk (see Fig. \ref{disk}). This form reminds us of the domain-wall solution considered in the previous case \cite{instanton1}. Thus the present solution preserves the analogy with the picture of the Skyrmion trapped on the domain wall \cite{Atiah-Manton}.

Even more interesting is the behavior of the optimal coordinates of the solution, which describe the position of the domain wall in the $z$ direction and the size of the instanton. We find, that for large values of $\sigma$ the optimal position of the instanton is located somewhere between IR and UV boundaries of the bulk space. This is in contrast with the usual behavior of the instanton solutions in $AdS$ \cite{Sakai-baryon, Wulzer}, which tend to fall on the IR boundary. In our treatment we find, that the chiral condensate provides a counter force, which drives the solution from IR boundary into the bulk. This result is directly related to the fact that the instantons on the $D8$ branes prevent the F1 string to shrink in the Sakai-Sugimoto picture  discussed above. Our treatment shows that the chiral condensate and the related bifundamental field $X$ play the important role of stabilizing the position and the size of the instanton solution, which we consider in this work. This points to the fundamental 
relation between baryon and chiral physics in QCD.

\begin{figure}[h!]
\includegraphics[width=0.5 \linewidth]{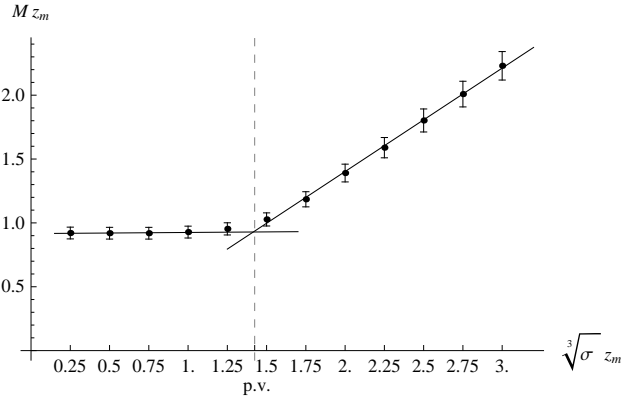}
\caption{\label{masses} The dependence of the baryon mass on the chiral condensate. The lines are constructed by the least squares fit. The fit is well inside the 5\% precision error bars of the energy values. The dashed line represents the physical values (p.v.) of the parameters, which provide the fit of the hard wall model to the real observables.}
\end{figure}

The fact that the chiral condensate governs the $z$ position of the solution and the energy of the solution that is the mass of the  baryon  suggests to study the dependence of the baryon mass on the chiral condensate in our model. On Fig. \ref{masses} we plot the energy (mass) of the optimal solution with respect to the cube root of $\sigma$, everything normalized to the position of the IR boundary $z_m$. We see  that for small $\sigma$ the energy does not change since the solution lyes just on the IR boundary, but for larger $\sigma$ the solution detaches from the boundary and the mass start to grow with the chiral condensate. The striking feature of this plot is the obviously linear dependence of the baryon mass on the energy scale of the condensate. This fits with the Ioffe's formula with the additional constant term
\cite{Ioffe}
\begin{equation}
\label{result}
M_b = \mathrm{Max}\big[(a \la \bar{q} q \ra^{1/3} + b), \ c \big].
\end{equation}
We find  the approximate value  $a \approx 1.72$, $b \approx - 70 Mev$, $c \approx 300 Mev$, where the normalization (\ref{sigma}) is used. This formula underestimates the mass of the physical baryon by the factor of 3 and this effect may be attributed to the uncertainties in the parameter values in the ``hard-wall'' model or to the oversimplification of the bulk metric and the $N_f=2$ approximation. Nevertheless, we stress the linear dependence observed in (\ref{result}), which is in qualitative agreement with the relations (\ref{Form_Ioffe}), (\ref{linear}). It provides the very interesting resolution
of the longstanding puzzle concerning the origin of the Ioffe's formula. And points out the important interplay between the chiral and baryonic physics in QCD.

\section{Conclusion} 
The dyonic instanton realization of the baryon provides us with a new tool to take into account the chiral symmetry
breaking effects in the baryon physics. Certainly it is very important in many aspects and much more detailed investigation of the generalized Skyrmion is required beyond the oversimplified hard-wall model. However even within the hard-wall model one could answer an interesting questions. For instance one could consider the formfactors of the generalized Skyrmion, study the spacial variations of the chiral condensate value inside the solution and look for the geometry of the solutions with higher baryon charges.

\bibliographystyle{aipproc}   

\end{document}